\documentclass[reprint, amsmath,amssymb,aps]{revtex4-1}
\usepackage{hyperref}
\usepackage{graphicx}
\usepackage{epstopdf}
\usepackage{mathrsfs}
\usepackage{amsmath,amssymb,amsfonts,amsthm,latexsym,stmaryrd,mathtools,underscore,float}
\usepackage{dsfont}
\usepackage{esint}
\usepackage[usenames, dvipsnames]{color}
\usepackage{subfig}
\def\nn{\nonumber}
\def\be{\begin{equation}}
\def\ee{\end{equation}}
\def\ba{\begin{eqnarray}}
\def\ea{\end{eqnarray}}
\allowdisplaybreaks[4]
\begin{document}

\title{Comment on ``Attractor solutions in scalar-field cosmology" and ``How many \emph{e}-folds should we expect from high-scale inflation?"}

\author{Yu Han}
 \email{hanyu@xynu.edu.cn}
 \affiliation{College of Physics and Electrical Engineering, Xinyang Normal University, 464000 Xinyang, China}%Lines break automatically or can be forced with \\
\author{Long Chen}
 \email{\text{chenlong\_phy}@aliyun.com}
 \affiliation{College of Physics and Electrical Engineering, Xinyang Normal University, 464000 Xinyang, China}%Lines break automatically or can be forced with \\

\begin{abstract}	
In Ref.~\cite{Remmen:2013}, it was claimed  that in the spatially flat cosmological case there exists a unique conserved measure (up to normalization) on the  $(\phi,\dot{\phi})$ phase space for scalar field with $m^2\phi^2$ potential by finding a unique solution to the differential equation $(44)$ (in Ref.~\cite{Remmen:2013}) in the low-energy regime. In Ref.~\cite{Remmen:2014}, it was also claimed that a unique solution to the same differential equation was found in the high-energy regime and using this solution the authors calculated the expected total number of $\emph{e}$-folds of inflation. In this comment, we reanalyze the differential equation $(44)$ and obtain general solutions both in the low-energy and high-energy regime, which can include the solution in Ref.~\cite{Remmen:2013} and the solution in Ref.~\cite{Remmen:2014} as a special case in the corresponding energy regime. In this way, we find that following the constructions in Ref.~\cite{Remmen:2013} there actually exist infinitely many nonequivalent conserved measures for the scalar-field cosmology with $m^2\phi^2$ potential on the $(\phi,\dot{\phi})$ phase space. Moreover, through specific calculations, we also show that different choices of measures can lead to quite different predictions of the expected total number of $\emph{e}$-folds of inflation.

\end{abstract}

\maketitle

%\tableofcontents
\section{Differential equation for Carroll-Remmen measure}
On spatially flat Friedmann-Robterson-Walker background, the cosmological dynamics of general relativity with scalar-field potential $V(\phi)=m^2\phi^2$ can be studied in the two dimensional phase space $(\phi,\dot{\phi})$ in which $\dot{\phi}\equiv\frac{d\phi}{dt}$ and $t$ is the proper time. The Hamiltonian flow vector field is given by $v^a=\left(\frac{\partial}{\partial t}\right)^a$. Denoting $x\equiv\phi,y\equiv\dot{\phi}$,
the measure on phase space $\left(x,y\right)$  is a two-form $\boldsymbol{\sigma}_{ab}$
that can be written as
\be
\boldsymbol{\sigma}_{ab}=f\left(x,y\right)(\mathrm{d}x)_a\wedge(\mathrm{d}y)_b\label{sigmaxy}
\ee
for some function $f$. Conservation of the measure along $v^a$ is provided by $\mathcal{L}_{v}\boldsymbol{\sigma}_{ab}=0$, which can be expressed as
\begin{equation}
\nabla_a\left(fv^a\right)=0.\label{div(fv)}
\end{equation}

\begin{figure}[h]
\centering
\includegraphics[height=4.5cm,width=5cm]{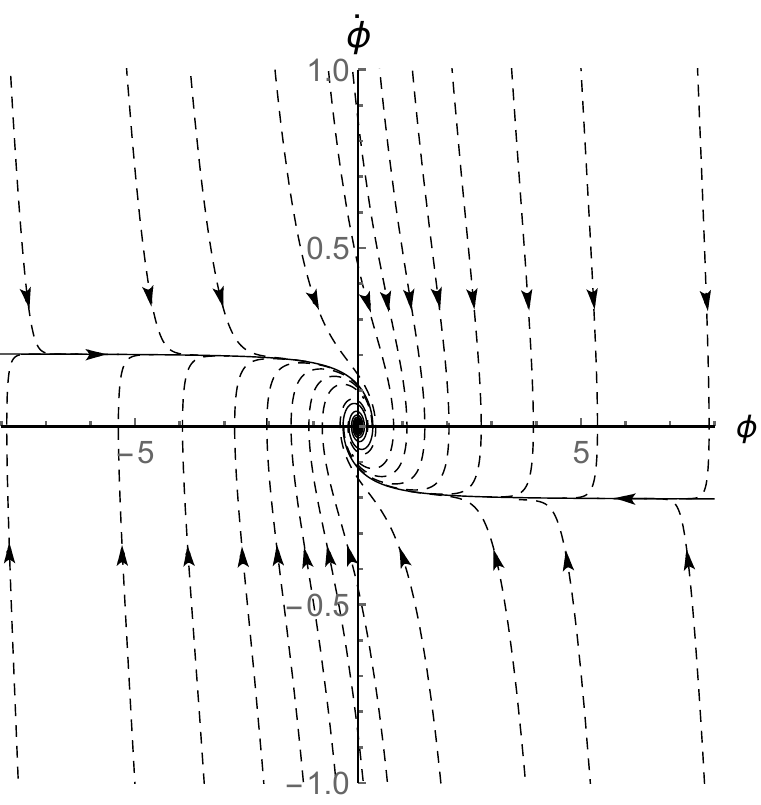}
\captionsetup{justification=raggedright}
\caption{Solutions for the Klein-Gordon equation $\ddot{\phi}+\sqrt{\frac{3\kappa}{2}}\sqrt{\dot{\phi}^2+m^2\phi^2}\dot{\phi}+m^2\phi=0$. Solid lines denote the attractive separatrices which approach $\dot{\phi}=\pm\sqrt{\frac{2}{3\kappa}}m$ at large field values. Plots are in $(\phi,\dot{\phi})$ space, in units where $\kappa=1$ and $m=0.25M_{\text{Pl}}$. }
\label{FIG1}
\end{figure}

As seen from Fig.~(\ref{FIG1}),  to describe the dynamics, it is convenient to reparametrize the vector
field in terms of polar coordinates $\left(r,\theta\right)$ defined by
\ba
r&=&\sqrt{y^2+m^2x^2}, \nn\\
x&=&\phi=\frac{r}{m}\cos\theta,\qquad y=\dot{\phi}=r\sin\theta.
\ea
Then, the measure can be written as
\be
\boldsymbol{\sigma}_{ab}=\frac{fr}{m}(\mathrm{d}r)_a\wedge(\mathrm{d}\theta)_b.\label{sigmadrdtheta}
\ee

Using the the equations
\ba
\frac{\partial r}{\partial t}&=&-3Hr\sin^2\theta,\label{drdt}\\
\frac{\partial \theta}{\partial t}&=&-\left(m+3H\sin\theta\cos\theta\right).\label{dthetadt}
\ea
in which $H$ is the Hubble parameter and $\dot{r}\equiv\frac{\mathrm{d}r}{\mathrm{d}t}$.
The Hamiltonian flow vector field can be expressed as
\be
\small
\left(\frac{\partial}{\partial t}\right)^a=
-3rH\sin^2\theta\left(\frac{\partial}{\partial r}\right)^a-\left(m+3H\sin\theta\cos\theta\right)
\left(\frac{\partial}{\partial \theta}\right)^a.\label{vartheta}\ee

Thus, Eq.~(\ref{div(fv)}) can be expressed as
\be
\frac{3\sin^2\theta}{r}\frac{\left(\partial Hr^2f\right)}{\partial r}+
\frac{\partial \left[f\left(m+3H\sin\theta\cos\theta\right)\right]}{\partial \theta}=0.\label{diffeq1}
\ee
Using the Friedmann equation
\ba
H^2=\frac{\kappa}{6}r^2\label{Friedmann}
\ea
where $\kappa=8\pi G$ and noticing that in an expanding Universe $H=\sqrt{\frac{\kappa}{6}}r$,
Eq. (\ref{diffeq1}) becomes
\be
\frac{\sin^2\theta}{r}\frac{\partial \left(r^3f\right)}{\partial r}+
\frac{\partial \left(\sqrt{\frac{2}{3\kappa}}mf+r\sin\theta\cos\theta f\right)}{\partial \theta}=0,\label{diffeq2}
\ee
which is the differential equation for Carroll-Remmen measure (see Eq.~(44) in Ref.~\cite{Remmen:2013}).

In Refs.~\cite{Remmen:2013} and ~\cite{Remmen:2014}, the authors claimed to find a unique solution to Eq. (\ref{diffeq2}) in the low-energy regime  $\sqrt{\kappa}r\ll m$ and a unique solution in the high-energy regime $\sqrt{\kappa}r\gg m$ respectively. However, as we show in the following sections, the solutions of Eq. (\ref{diffeq2}) are actually far from unique both in the low-energy regime and the high-energy regime.

\section{General solution to Eq. (\ref{diffeq2}) in the low-energy regime}

In the low-energy regime where $\sqrt{\frac{2}{3\kappa}}\frac{m}{r}\gg 1$, Eq.~(\ref{diffeq2}) is approximated by
\be
\frac{\sin^2\theta}{r}\frac{\partial \left(r^3f\right)}{\partial r}+
\sqrt{\frac{2}{3\kappa}}m\frac{\partial f}{\partial \theta}=0,\label{diffeq3}
\ee
which has the solution
\be
f= \frac{\Psi\left(\alpha\right)}{r^3}, \label{approf}
\ee
in which
\be
\alpha\equiv\theta-\frac{1}{2}\sin(2\theta)+2\sqrt{\frac{2}{3\kappa}}\frac{m}{r}\label{alpha}
\ee
and $\Psi\left(\alpha\right)$ denotes an arbitrary  function of the variable $\alpha$. Moreover, since Eq.~(\ref{diffeq2}) remains unchanged under the transformation $\theta\rightarrow\theta+\pi$ (or $(x,y)\rightarrow(-x,-y)$), it implies that its solution should also be symmetric under such transformation. Hence, $\Psi(\alpha)$ should be a periodic function with period $\pi$.
Using Eq.~(\ref{sigmadrdtheta}), the measure becomes
\be
\boldsymbol{\sigma}_{ab}=\frac{\Psi(\alpha)}{mr^2}(\mathrm{d}r)_a\wedge(\mathrm{d}\theta)_b.
\label{sigmartheta}
\ee

There is the other way to obtain the asymptotic solution in  Eq.~(\ref{approf}). We parameterize the trajectories in the phase space $(\phi,\dot\phi)$  using the variables $(\beta,t)$ such that $\beta$ is invariant along each trajectory, i.e.,$\mathcal{L}_{v}\beta=0$. According to $\mathcal{L}_{v}\boldsymbol{\sigma}_{ab}=0$, the measure $\boldsymbol{\sigma}_{ab}$ should satisfy
\be
\boldsymbol{\sigma}_{ab}\propto(\mathrm{d}\beta)_a\wedge(\mathrm{d}t)_b.
\ee
$\beta$ can be expressed as a function of $r$ and $\theta$, which satisfies
\be
\frac{\mathrm{d}\beta}{\mathrm{d}t}
=\frac{\partial r}{\partial t}\frac{\partial \beta}{\partial r}+\frac{\partial \theta}{\partial t}\frac{\partial \beta}{\partial \theta}=0.
\ee
It is not difficult to check that in the low-energy limit $\frac{\mathrm{d}\alpha}{\mathrm{d}t}=0$, hence,  we find that in the low-energy limit $\beta=\Phi(\alpha)$ where $\Phi(\alpha)$ is also an arbitrary function of $\alpha$, thus we have
\ba
(\mathrm{d}\beta)_a\wedge(\mathrm{d}t)_b&\propto&\Phi'(\alpha)
\left(\frac{\partial \alpha}{\partial r}\frac{\partial t}{\partial \theta}-\frac{\partial \alpha}{\partial \theta}\frac{\partial t}{\partial r}\right)(\mathrm{d}r)_a\wedge(\mathrm{d}\theta)_b\nn\\
&=&\frac{4\sqrt{2}\Phi'(\alpha)}{\sqrt{3\kappa}r^2}(\mathrm{d}r)_a\wedge(\mathrm{d}\theta)_b.\label{dbetadtdrdtheta}
\ea
 Eq.~(\ref{dbetadtdrdtheta}) is identical with Eq.~(\ref{sigmartheta}) if we require $\Phi'(\alpha)\propto\Psi(\alpha)$.

Using (\ref{approf}), we obtain the general solution of $f$ to Eq.~(\ref{diffeq2}) in the low-energy regime,
\ba
f=\frac{1}{r^3}
&&\Bigg[\Psi\left(\alpha\right)\left(1-\sqrt{\frac{3\kappa}{2}}\frac{r}{m}\sin\theta\cos\theta\right)\nn\\
&&~-\frac{1}{2}\sqrt{\frac{3\kappa}{2}}\frac{r}{m}\sin^4\theta\Sigma\left(\alpha\right)+\mathcal{O}\Bigg],\label{solf}
\ea
in which $\Sigma\left(\alpha\right)\equiv\frac{\mathrm{d}\Psi\left(\alpha\right)}{\mathrm{d}\alpha}$ and $\mathcal{O}$ is a function containing functions of $\theta$ and higher order terms of $\sqrt{\frac{3\kappa}{2}}\frac{r}{m}$.

In Ref.~\cite{Remmen:2013}, the authors assumed that in the low-energy limit $\frac{\partial f}{\partial \theta}=0$, using this assumption, they find a solution of Eq. (\ref{diffeq3}),
\be
f\rightarrow \frac{A}{r^3},\label{appsolR}
\ee
where $A\in\mathbb{R}$. Using this solution, they claimed to obtain a unique solution of Eq. (\ref{diffeq2})
 (see Eq.~(53) in Ref.~\cite{Remmen:2013}):
\be
f= \frac{A}{r^3}\left(1-\sqrt{\frac{3\kappa}{2}}\frac{r}{m}\sin\theta\cos\theta+\mathcal{O}\right).\label{solR}
\ee
Then, according to the Cauchy-Kowalevski theorem \cite{Kowalevski:2009,Nandakumaran:2020}, the authors concluded that there exists a unique measure (up to normalization) for the scalar-field cosmology with $m^2\phi^2$ potential.

Comparing the solution (\ref{solR}) with our solution in Eq. (\ref{solf}), we find that when $\Psi(\alpha)=A$  the solution in Eq. (\ref{solf}) can reproduce the solution in (\ref{solR}).

However, the assumption $\frac{\partial f}{\partial \theta}=0$ in the low-energy limit is too restrictive. To observe this, we rewrite Eq.~(\ref{diffeq3}) as
\be
\left(\sqrt{\frac{3\kappa}{2}}\frac{r}{m}\sin^2\theta\right)r\frac{\partial (r^3f)}{\partial r}+\frac{\partial (r^3f)}{\partial \theta}=0.\label{diffeq3b}
\ee
In the low-energy regime, although the prefactor $\sqrt{\frac{3\kappa}{2}}\frac{r}{m}\sin^2\theta\ll1$, $r\frac{\partial (r^3f)}{\partial r}$ may be large, causing the first term comparable to the second term on the left hand side of Eq.~(\ref{diffeq3b}) and hence neither of these two terms should be easily discarded. Indeed, as we see from Eq.~(\ref{approf}), any $\Psi(\alpha)$ with $\Psi(\alpha)\neq \text{constant}$ can easily satisfy this condition.   Due to the huge ambiguities in choosing $\Psi(\alpha)$,  there actually exist infinitely many solutions to the differential equation (\ref{diffeq2}) in the low-energy regime.

A conserved measure $\boldsymbol{\omega}_a$ on the space of trajectories can be constructed from the phase space measure by defining $\boldsymbol{\omega}_a\equiv\boldsymbol{\sigma}_{ab}v^b$, then obviously we have $\mathcal{L}_{v}\boldsymbol{\omega}_{a}=0$,
and it is not difficult to show that on any surface transverse to those trajectories
\be
\int_{\mathcal{S}}\boldsymbol{\omega}_{a}=\int_{\mathcal{S}'}\boldsymbol{\omega}_{a}=\int_{\mathcal{S}}\boldsymbol{\tilde{\omega}}_{a}
=\int_{\mathcal{S}'}\boldsymbol{\tilde{\omega}}'_{a},\label{promeasure}
\ee
in which the surfaces $\mathcal{S}$ and $\mathcal{S}'$ are passed through by the same bunch of trajectories only once, and measure $\boldsymbol{\tilde{\omega}}_{a}$ and $\boldsymbol{\tilde{\omega}}'_{a}$ represent respectively the restriction of $\boldsymbol{\omega}_a$ on  $\mathcal{S}$ and $\mathcal{S}'$.

Considering two concentric circles $\mathcal{C}$ and $\mathcal{C}'$ (both centered at the origin of the  space $(mx,y)$) with radius $r$ and $r'$ respectively, since all of the phase space orbits pass through $\mathcal{C}$ or $\mathcal{C}'$ only once, using  Eq.~(\ref{promeasure}) and (\ref{sigmadrdtheta}), we have
\be
\oint_{\mathcal{C}'}\boldsymbol{\tilde{\omega}'}_{a}=
\oint_{\mathcal{C}}\boldsymbol{\tilde{\omega}}_{a}
=\int_{0}^{2\pi}\frac{1}{m}\sqrt{\frac{3\kappa}{2}}fr^3\sin^2\theta d\theta.\label{inteC}
\ee
With suitable normalization constant $B$, we define
\be
P(\theta,r)\equiv \frac{B}{m}\sqrt{\frac{3\kappa}{2}}fr^3\sin^2\theta \label{pdf}
\ee
as the probability distribution function, satisfying 
\be
\int_{0}^{2\pi}P(\theta,r) d\theta=1.\label{intP}
\ee
 Using Eq.~(\ref{approf}), we find that in the low-energy limit the integral becomes
\be
\int_{0}^{2\pi}P(\theta,r) d\theta=\sqrt{\frac{3\kappa}{2}}\frac{B}{m}\int_{0}^{\pi}\Psi(\alpha) d\alpha,\label{intPsi}
\ee
therefore, to obtain a well defined probability distribution function, $\Psi(\alpha)$ should  be a positive definite function and the integral of $\Psi(\alpha)$ over a period must be finite.

\begin{figure}[h]
\centering
\includegraphics[height=4.5cm,width=5cm]{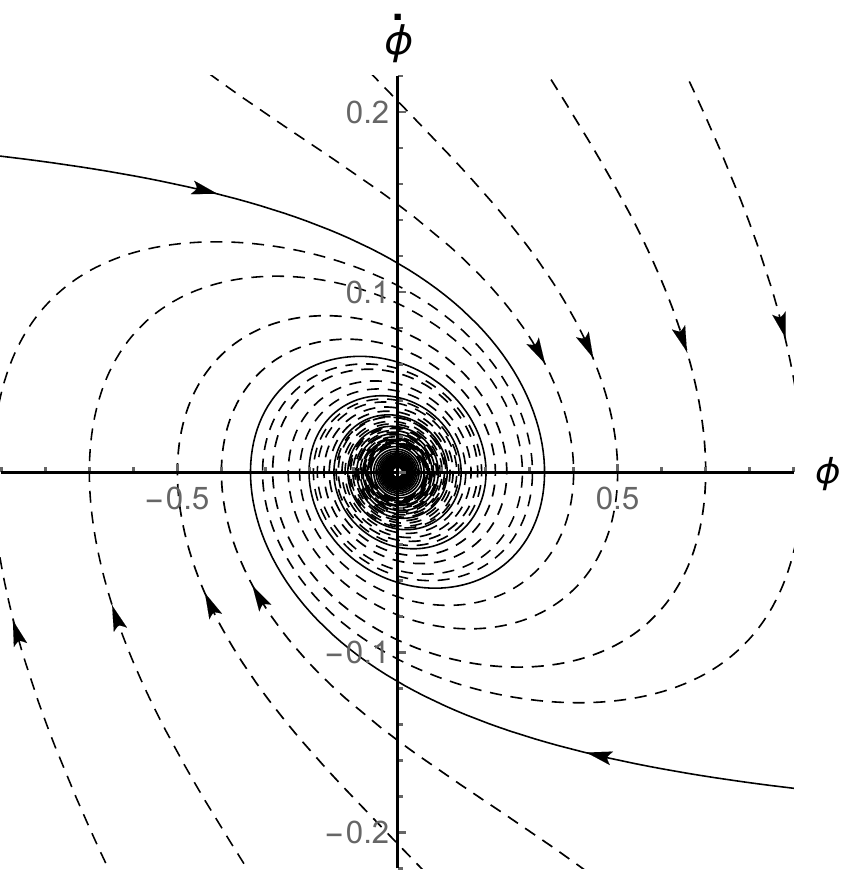}
\captionsetup{justification=raggedright}
\caption{Dynamical behavior of scalar-field cosmology with $m^2\phi^2$ potential near the origin of $(\phi,\dot{\phi})$ space, in units where $\kappa=1$ and $m=0.25M_{\text{Pl}}$. }
\label{FIG2}
\end{figure}

It is worth remarking that at the origin $\phi=\dot{\phi}=0$ the asymptotic solution (\ref{approf}) diverges. This divergence results from the fact that all trajectories converge at the origin (see Fig.~(\ref{FIG2}) ) and can be explained from the perspective of the probability distribution function, from Eq. (\ref{intP}) we find
\be
\int_{0}^{2\pi}f\sin^2\theta  d\theta=\sqrt{\frac{2}{3\kappa}}\frac{m}{B}\frac{1}{r^3}.\label{intf}
\ee
 It is easy to see that the function $f$ diverges as $r\rightarrow 0$. 
 Therefore, we conclude that there exist infinitely many conserved measures on the $(\phi,\dot{\phi})$ phase space for scalar field with $m^2\phi^2$ potential, excluding the origin.

Another explanation of the divergence of $f$ at the origin from a slightly different perspective can be found in Ref. \cite{Remmen:2013} (see the paragraph below Eq. (44) of Ref. \cite{Remmen:2013} ).
\section{General solution to Eq. (\ref{diffeq2}) in the high-energy regime}
In the high-energy regime where $\sqrt{\frac{2}{3\kappa}}\frac{m}{r}\ll 1$, Eq.~(\ref{diffeq2}) is approximated by
\be
\frac{\sin^2\theta}{r^2}\frac{\partial \left(r^3f\right)}{\partial r}+
\frac{\partial \left(\sin\theta\cos\theta f\right)}{\partial \theta}=0.\label{diffeq4}
\ee
Under the transformation $\theta\rightarrow\theta+\pi$ (or $(x,y)\rightarrow(-x,-y)$), the solution should be symmetric.
The general solution to Eq.~(\ref{diffeq4}) reads
\be
f=\frac{\Theta\left(\text{sgn}(y)r\cos\theta\right)}{\text{sgn}(y)r^2\sin\theta}
=\frac{\Theta\left(\text{sgn}(y)mx\right)}{|y|\sqrt{y^2+m^2x^2}},\label{approf2}
\ee
in which $\Theta(\text{sgn}(y)r\cos\theta)$ is an arbitrary function of $\text{sgn}(y)r\cos\theta$, and this solution is obviously symmetric under $\theta\rightarrow\theta+\pi$ (or $(x,y)\rightarrow(-x,-y)$).
Moreover, to obtain a well defined probability distribution function, $\Theta$ requires to be a positive definite function and using Eq. (\ref{intf}) we find $\Theta$ must approach zero as $r\rightarrow\infty$.

 By  expanding $\Theta\left(\text{sgn}(y)r\cos\theta\right)$ into a power series of $\text{sgn}(y)r\cos\theta$:
\be
\Theta\left(\text{sgn}(y)r\cos\theta\right)=\sum_{i}C_{i}\text{sgn}^i(y)r^{i}\cos^{i}\theta
\ee
where $i$, $C_{i}\in\mathbb{R}$, the solution in Eq.~(\ref{approf2}) becomes
\be
f=\sum_{i}\frac{C_{i}\text{sgn}^i(y)r^{i}\cos^{i}\theta}{r^2|\sin\theta|},\label{approf2b}
\ee
which is slightly different from Eq.~(47) of  Ref.~{\cite{Remmen:2014}}:
\be
f=\sum_{i}\frac{C_{i}r^{i}|\cos^{i}\theta|}{r^2|\sin\theta|}.\label{approf2bR}
\ee
 Actually, the solution (\ref{approf2bR}) is too restrictive because an additional symmetry which is absent in Eq.~(\ref{diffeq4})---the symmetry under $(x,y)\rightarrow(-x,y)$---is enforced on the solution.

In Ref.~{\cite{Remmen:2014}},  by imposing the requirements that at fixed $\theta$  the solution in Eq.~(\ref{approf2bR}) should be a nonincreasing function of $r$ and infinitely differentiable everywhere (except  $\sin\theta=0$) in the large $r$ case, the authors argued that $i$ should be $2$, which selects
\be
f\propto\frac{\cos^{2}\theta}{|\sin\theta|}\label{approf2R}
\ee
as the unique physical solution in the high energy regime (see the paragraph before Eq.~(48) of  Ref.~{\cite{Remmen:2014}}).
However, the arguments there were incorrect because the terms with higher power exponents of $i$ in the solution (\ref{approf2bR}) do not necessarily imply that  $f$  must be an increasing function of $r$ at fixed $\theta$ in the large $r$ regime. As a counterexample,  we can set
\be
f=\frac{1}{r^2|\sin\theta|}\exp\left[-\frac{1}{\gamma^2}\left(\frac{\sqrt{\kappa}}{m}r\cos\theta\right)^2\right],\label{cexample}
\ee
which can be expanded as
\be
f=\frac{1}{|\sin\theta|}\sum_{i=0}^{\infty}\frac{(-\kappa)^i}{i!(\gamma m)^{2i}}r^{2(i-1)}\cos^{2i}\theta,
\ee
 in which $\gamma$ is an arbitrary constant. It is easy to verify that $f$ in Eq. (\ref{cexample}) is a solution of Eq. (\ref{diffeq4}). This expansion contains infinite terms with higher power exponents of $i$ but it is still a nonincreasing function of $r$ at fixed $\theta$ and also infinitely differentiable everywhere (except $\sin\theta=0$) in the large $r$ regime.

The normalized probability distribution function corresponding to $f$ in Eq. (\ref{cexample}) is given by
\be
P(\theta,r)=\frac{\sqrt{\kappa }}{2\sqrt{\pi}|\gamma|m}r|\sin \theta|\exp\left[-\frac{1}{\gamma^2}\left(\frac{\sqrt{\kappa}}{m}r\cos\theta\right)^2\right].\label{Pcexample}
\ee
The probability distribution function can be used to calculate the expected total number of $\emph{e}$-folds of inflation. Using the result in Ref. \cite{Remmen:2014}, for the scalar-field potential $m^2\phi^2$,  the total number of $\emph{e}$-folds that a trajectory undergoes is given by
\be
N_{\text{tot}}(\Lambda_{\text{UV}},\theta)=\frac{3}{2}\left(\frac{\Lambda_{\text{UV}}}{m}\right)^2\cos^2 \theta,
\ee
in which $\Lambda_{\text{UV}}$ is the ultraviolet cutoff and $\theta$ is the angle that  each trajectory starts
at on the surface where $H=\sqrt{\frac{\kappa}{6}}r=\Lambda_{\text{UV}}$.

The expected total number of $\emph{e}$-folds of inflation is  given by
\ba
\langle N_{\text{tot}}\rangle
&=&\int_{0}^{2\pi}N_{\text{tot}}(\Lambda_{\text{UV}},\theta)P(\theta,r)\Big{|}_{\sqrt{\kappa}r=\sqrt{6}\Lambda_{\text{UV}}} d\theta \nn\\
&=&\frac{\gamma^2}{8}.\label{expN}
\ea
Surprisingly, the final result is independent of the ultraviolet cutoff and the inflaton mass. From this result, we see that to obtain the required number of e-folds needed to solve the horizon problem we must have $\gamma>20$, which is completely different from the predictions given in Ref.~\cite{Remmen:2014} by choosing $f$ in Eq. (\ref{approf2R}), therefore, we can conclude that different choices of measures can lead to completely different predictions of the expected total number of $\emph{e}$-folds of inflation.

\begin{figure}[h]
\centering
\includegraphics[height=4cm]{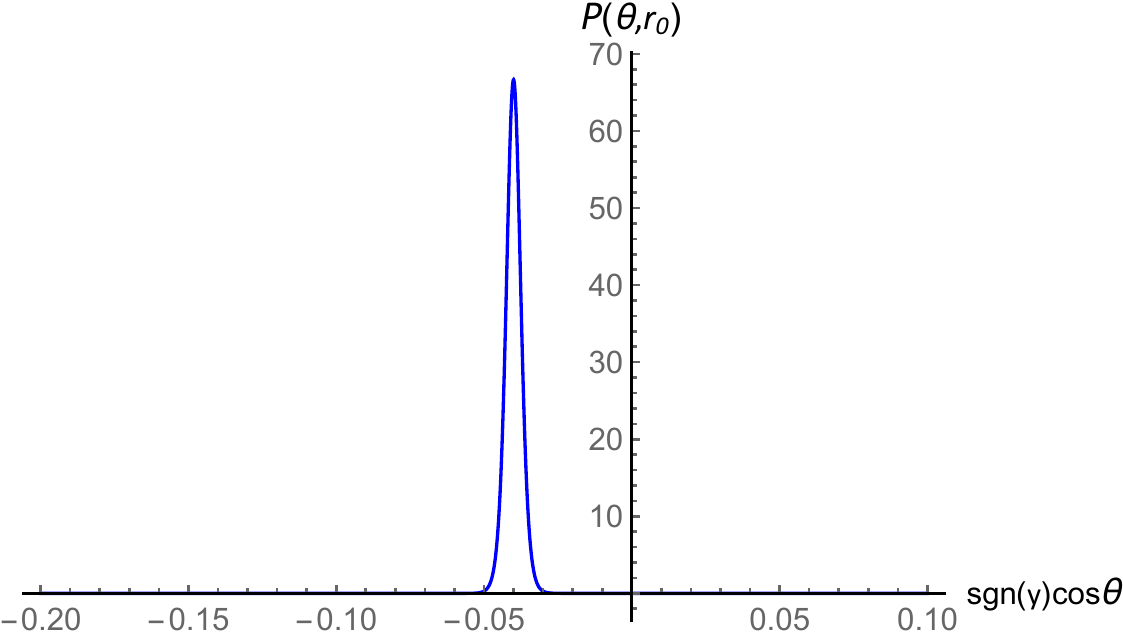}
\captionsetup{justification=raggedright}
\caption{The probability distribution function $P(\theta,r_0)$ at $\sqrt{\kappa}r_{0}/m=10^3$ with $m=10^{-2}M_{\text{Pl}}$.}
\label{FIG3}
\end{figure}

In Fig.~(\ref{FIG3}), we show the numerical result for the probability distribution function $P(\theta,r_0)$ at $\sqrt{\kappa}r_0/m=10^3$ by choosing the initial probability distribution function using $f$ in  Eq. (\ref{solR}) at $\sqrt{\kappa}r/m=0.1$ . The numerical result shows that the probability distribution function is extremely sharply peaked near $\cos\theta=0$ in the high energy regime, which is obviously inconsistent with the result in Eq.~(\ref{approf2R}).

\begin{figure}[h]
\centering
\includegraphics[height=4cm]{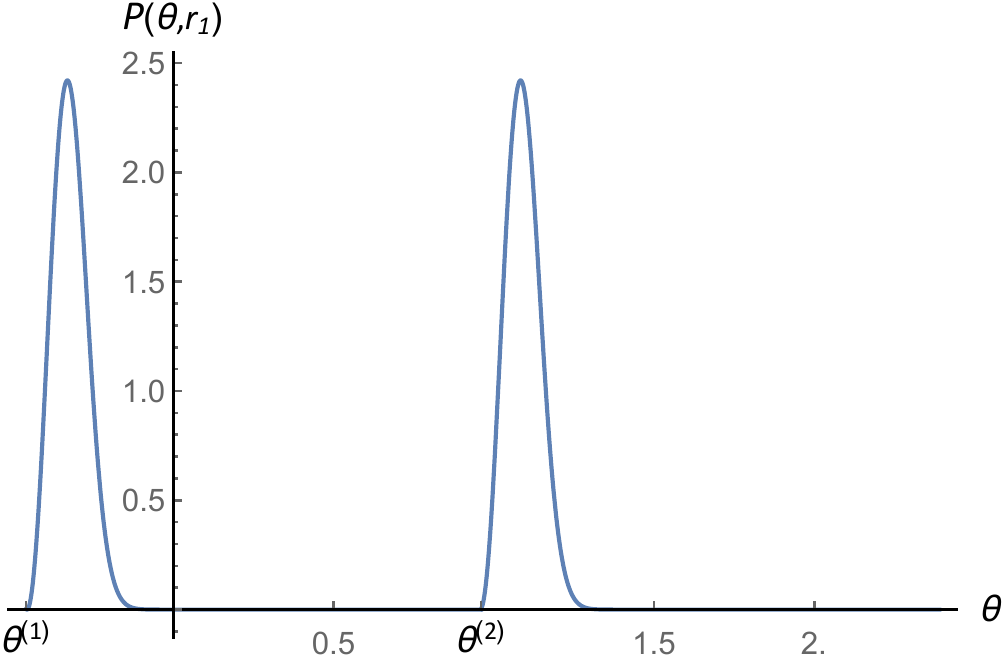}
\captionsetup{justification=raggedright}
\caption{The probability distribution function $P(\theta,r_1)$ at $\sqrt{\kappa}r_{1}/m=0.1$. $\theta^{(1)}$ and $\theta^{(2)}=\theta^{(1)}+\pi$ are the angles at which the two attractive separatrices intersect the circle $r=r_1$ respectively in the space $(mx,y)$.}
\label{FIG4}
\end{figure}

 On the other hand, if we choose the initial probability distribution function using $f$  in  Eq.~(\ref{approf2R}) at $\sqrt{\kappa}r/m=10^3$, the numerical result  shows that the probability distribution function $P(\theta,r_1)$  at $\sqrt{\kappa}r_1/m=0.1$ is also sharply peaked in certain direction, i.e., $\frac{\partial f}{\partial \theta}$ can be very large at certain angle in the low-energy regime (see Fig.~(\ref{FIG4})),  which is obviously inconsistent with the result in Eq.~(\ref{solR}).

\section{Conclusion}
  In Ref.~\cite{Remmen:2013} and Ref.~\cite{Remmen:2014}, the authors claimed to find a unique solution to the differential equation (44) in Ref.~\cite{Remmen:2013} (which is Eq. (\ref{diffeq2}) in this comment) in the low-energy and high energy regime respectively, hence, they conclude that there exists a unique conserved measure (up to normalization) on the $(\phi,\dot{\phi})$ phase space  for scalar field with $m^2\phi^2$ potential. Moreover, in Ref.~\cite{Remmen:2014}, the authors also used the solution they obtained to calculate the expected total number of $\emph{e}$-folds of inflation. However, in this comment, we reanalyze the solutions of the same differential equation and obtain quite different results.
  
  First, we find the asymptotic solution (\ref{approf}) to the differential equation (\ref{diffeq2}) in the low-energy limit using two approaches, using the asymptotic solution we obtain the general solution in Eq. (\ref{solf}) in the low-energy regime.  We find that the ``unique" solution in Eq. (\ref{solR}) (obtained in Ref.~\cite{Remmen:2013}) is a special case of the general solution in Eq. (\ref{solf}), corresponding to $\Psi(\alpha)=\text{constant}$. Due to the huge ambiguities in choosing $\Psi(\alpha)$, we conclude that following the constructions in Ref.~\cite{Remmen:2013} there exist infinitely many nonequivalent conserved measures for the scalar field with $m^2\phi^2$ potential on the $(\phi,\dot{\phi})$ phase space in the low-energy regime.
  
  Next, we obtain the general solution (\ref{approf2}) to the differential equation (\ref{diffeq2}) in the high energy regime, which also includes the ``unique" solution in Eq. (\ref{approf2R}) (obtained in Ref.~\cite{Remmen:2014}) as a special case. Moreover, through specific calculations, we show that choosing a  measure different from (\ref{approf2R}) can lead to completely different predictions of the expected total number of $\emph{e}$-folds of inflation.
  
  In addition, using the numerical results in Fig.~(\ref{FIG3}) and (\ref{FIG4}), we also show that the behavior of the solution in Eq. (\ref{solR}) in the high-energy regime are completely different from the solution in Eq. (\ref{approf2R}), and the behavior of the solution in Eq. (\ref{approf2R}) in the low-energy regime are completely different from the solution in Eq. (\ref{solR}), which imply that the claims of the solution in Eq. (\ref{solR}) to be the unique solution in the low-energy regime and the solution in Eq. (\ref{approf2R}) to be the unique solution in the high-energy regime  are  inconsistent with each other.

\begin{acknowledgements}
 This work is supported by NSFC (Grant No.11905178) and key scientific research projects in universities of Henan Province (Grant No.25A140014) and Nanhu Scholars Program for Young Scholars of Xinyang Normal University.
\end{acknowledgements}

\end{document}